\documentclass[12pt,preprint]{aastex}
\newcommand\etal    {{et~al.}~}

\usepackage{psfig}
\usepackage{epsf}
\usepackage{flushrt}
\usepackage{graphicx}

\begin{document}

\title{The Evolution of the Galaxy Cluster Luminosity-Temperature Relation}

\author{Megan C. Novicki}
\affil{University of Hawaii, Institute for Astronomy, 2680 Woodlawn Drive, Honolulu, HI, 96822 USA}

\author{Manuela Sornig}
\affil{Erzherzog-Johann Universitaet, Rechbauerstrasse 12, A-8010, Graz, Austria}

\author{J. Patrick Henry}
\affil{University of Hawaii, Institute for Astronomy, 2680 Woodlawn Drive, Honolulu, HI, 96822 USA}

\begin{abstract}
We analyzed the luminosity-temperature (L-T) relation for 2 samples of galaxy clusters which have all been observed by the {\em ASCA} satellite.  We used 32 high redshift clusters (0.3$<$z$<$0.6), 53 low redshift clusters (z$<$0.3), and also the combination of the low and high redshift datasets.  This is one of two surveys to use only {\em ASCA} data, and has the largest number of high redshift clusters.  We assumed a power law relation between the bolometric luminosity of the galaxy cluster and its integrated temperature (uncorrected for cooling flows) and redshift (L$_{bol,44}$=CT$^{\alpha}$(1+z)$^A$).  We found that for an $\Omega_{M}$=1.0 $\Omega_{\Lambda}$=0.0 universe, A = 1.134$^{+1.057}_{-1.073}$$\pm1.66$, $\alpha$ = 2.815$^{+0.322}_{-0.316}$$\pm0.42$, and log(C) = $-$1.167$^{+0.216}_{-0.221}$$\pm0.25$, and for $\Omega_{M}$=0.3 $\Omega_{\Lambda}$=0.7, A = 2.052$^{+1.073}_{-1.058}$$\pm1.63$, $\alpha$ = 2.822$^{+0.320}_{-0.323}$$\pm0.43$, and log(C) = $-$1.126$^{+0.223}_{-0.219}$$\pm0.26$ (all errors at 68\% confidence for one and two interesting parameters, respectively).  We found the dispersion at constant kT in this relation to be $\Delta$logL=0.282 for $\Omega_{M}$=1.0 $\Omega_{\Lambda}$=0.0, and $\Delta$logL=0.283 for $\Omega_{M}$=0.3 $\Omega_{\Lambda}$=0.7.

The results for the combined dataset and those found using the low and high redshift clusters are consistent, independent of cosmology, with previous estimates of L$\sim$T$^3$ found by other authors.  The observed weak or zero evolution agrees with the predictions of models that produce L$\sim$T$^3$ incorporating an initial source of non-gravitational energy before cluster collapse.
\end{abstract}

\section{Introduction} 

Galaxy clusters have three major mass components; the galaxies themselves comprise about 1\% of the cluster mass, about 10\% of the mass is contained in the hot ($\sim$10$^7$K) X-ray emitting gas of the intracluster medium (ICM), and the rest of the mass is contained in dark matter.  The dark matter itself probably exhibits self-similar scaling of its properties down to at the least the sizes of groups of galaxies due to its collisionless properties, and has been modeled with precision (e.g. Navarro, Frenk, \& White 1997).  The ICM does not have the same scaling properties as the dark matter.  Instead it has a core of hot matter which is more extended with respect to self-similar scaling in groups of galaxies and small galaxy clusters than in rich clusters (Ponman \etal 1999).

This break in the self-similar scaling between the hot intracluster medium and the dark matter raises questions about the formation of large structures.  The leading explanation for the break is the presence of additional non-gravitational energy in the early stages of galaxy cluster formation, proposed by Kaiser (1991) and Evrard \& Henry (1991).  This energy produces an initial excess entropy that has a stronger effect on smaller and cooler galaxy clusters than high-mass clusters if the excess entropy is relatively constant for all clusters.  There are five reasons to suspect an entropy excess.  The first is the difference between the chemical properties and the spatial distribution of the ICM in groups and clusters with ICM temperatures less than about 1keV and those with ICM temperatures greater than 1keV (Renzini 1999).  Second, the observed mass-temperature (M-T) relation is steeper than the relation predicted by self-similar scaling for clusters of ICM temperatures less than about 3keV (Horner, Mushotzky, \& Scharf 1999, Finoguenov \etal 2001).  Third, Ponman \etal (1999) found direct evidence of an entropy floor at low cluster temperature by looking at the entropy of the cluster gas at a fiducial radius as a function of the cluster temperature.  Fourth, entropy injection at early epochs is necessary to explain the correlation function and level of the X-ray background (Pen 1999; Wu, Fabian \& Nulsen 2000).

The fifth and final reason to suspect a substantial initial entropy is the relation between the bolometric luminosity of galaxy clusters and their temperature (L-T).  If the simple scaling laws were applicable, the luminosity would scale as $\sim$T$^2$ (Kaiser 1986).  The observed L-T relation is actually closer to L$\sim$T$^3$ (Henry \& Arnaud 1991, David \etal 1993, Henry 1997, Arnaud \& Evrard 1999, Borgani \etal 1999, Riechart \etal 1999, Fairley \etal 2000, Henry 2000).  Including an initial entropy in numerical simulations of galaxy cluster formation can produce an L-T relation closer to the observational value (Bialek \etal 2001, Tozzi \& Norman 2001).  

Initial entropy can be introduced into cluster formation scenarios at different times and from different sources such as star formation or active galactic nuclei (AGN).  If the entropy is not present when the gas is accreted onto the cluster, a much higher level is required to overcome the effects of cooling of high density gas in the cluster core.  Wu, Fabian \& Nulsen (2000) concluded that the energies required for the initial entropy may be provided by either supernovae or AGN, however, the conditions for which the supernova energy injection is large enough to produce the observed relations are highly contrived.  They found that AGN easily produce sufficient energy although the mechanism for using this energy to heat the hot gas is not known.  Fujita (2001) found that blast waves driven by quasars are a possible heating mechanism for the intragroup gas prior to falling into the galaxy clusters.  

Not all studies have concluded that additional entropy is required to produce the observed scaling laws of L-T and M-T.  Muanwong \etal (2001) simulated galaxy cluster formation including radiative cooling with cool gas dropout and were able to reproduce L$\sim$T$^3$ without adding any entropy to the gas.  Voit \& Bryan (2001) and Voit \etal (2002) propose a similar mechanism in which radiative cooling with subsequent supernova heating eliminate low-entropy gas.  In these models the self-similar scaling is broken because cooling is more efficient at lower temperatures.

The L-T relation has been well studied at low redshift, however, the situation at higher redshift is less clear. Self-similar scaling laws predict that L$\sim$T$^2$(1+z)$^{1.5}$ for $\Omega_M$=1 (where $\Omega_{M}$ is the present matter density in units of the critical density), which would indicate a strong positive evolution, but these simple scaling laws are contra-indicated by current observational evidence.  Henry \etal (1994), using three redshift bins, found that there is moderate to no evolution out to a redshift of 0.33.  Mushotzky \& Scharf (1997), using galaxy clusters that had been observed with {\em ASCA}, found evidence for no evolution out to a redshift of 0.4.  Fairley \etal (2000) found that L$\sim$T$^{3.15}$(1+z)$^{0.60\pm0.38}$ for an $\Omega_{M}$=0.3 universe, which is consistent with no evolution in the L-T relation out to a redshift of 0.8.  Sadat \etal (1999) found some evidence for positive evolution, although their analysis estimates $\Omega_{M}\sim$0.85, which is in disagreement with the majority of recent results which indicate an $\Omega_{M}$ of 0.2 to 0.4.   Arnaud, Aghanim, \& Neumann (2001) used {\em ASCA} and {\em ROSAT} data for 25 hot (kT$>3.5$keV) clusters, and found that A is positive for a flat universe with $\Omega_{M}$=0.4.  

The L-T relation not only plays an important role in discerning the physics behind the formation of galaxy clusters, as summarized above, but it provides a link between observations of clusters and estimation of cosmological parameters. Cluster number abundance evolution can constrain the shape and amplitude of the mass fluctuation power spectrum as well as the matter density of the universe.  When using luminosity to constrain cosmology, it is necessary to modify the mass-luminosity (M-L) relation using the observed L-T relation to incorporate the fact that the luminosities are not given by the self-similar scaling laws.  Also, galaxy cluster samples are selected via their X-ray luminosity, but the X-ray temperature is more directly linked to the cluster mass and hence models of cluster abundance.  The L-T relation is used to convert the luminosity selection function to the temperature selection function.  These topics are beyond the scope of this paper, but for a complete discussion of the methods, see Henry (2000).  

We constructed the L-T relation for a sample of galaxy clusters which have all been observed by {\em ASCA}.  The sample is comprised of all of the clusters of which we are aware with 0.3$<$z$<$0.6 that have been observed with {\em ASCA}, in combination with the clusters in the sample in David \etal 1993 for which there are {\em ASCA} data available, giving us a total sample of 85 clusters.  This sample is not flux limited but is assumed to be representative of galaxy clusters.  While data from {\em Chandra} or {\em XMM} might be better suited for our purposes, {\em ASCA} data are the only ones available for a sufficiently large number of clusters to conduct a statistically relevant uniform study.  The {\em ASCA} satellite has two high-quality spectroscopic instruments covering an energy range similar to the typical temperature range of galaxy clusters (kT$\sim$2$-$10keV).  Many of the previous studies have used {\em ROSAT} data alone or in combination with {\em ASCA} data.  {\em ROSAT} is only sensitive to a small range of energies (0.1$-$2.4keV) which are less than the average cluster temperature.  Using only {\em ASCA} data gave us a sample which is uniform and allows us to do high precision statistical analysis.  The sample most similar to ours is the {\em ASCA} Cluster Catalogue, which contains 273 galaxy clusters and groups all observed with {\em ASCA} (Horner 2001).  This work used both pointed and serendipitous observations of clusters and explored the L-T, M-T, and optical velocity dispersion ($\sigma$)-T relation out to a redshift of about 0.5.

We examined the slope of the L-T relation at low redshift ($<$0.3, 53 clusters), high redshift (0.3$<$z$<$0.6, 32 clusters), and for the dataset as a whole.  We investigated whether the L-T relation evolves with redshift for two different cosmologies thereby testing the prediction of little to no evolution out to z$\sim$1 if there was a significant source of non-gravitational energy available durning the early stages of cluster formation (Bialek \etal 2001, Tozzi \& Norman 2001, Bower \etal 2001).  We assumed a Hubble constant of 50 km s$^{-1}$Mpc$^{-1}$ throughout.

\section{Data}

For the z$\geq$0.3 clusters, we determined the integrated temperature, uncorrected for cooling flows, and luminosity of the high redshift clusters using {\em ftools} and {\em XSPEC}. Cooling flow corrections are generally not possible with {\em ASCA} for clusters at these redshifts because of the spatial resolution of the observations.  The {\em ASCA} Gas Imaging Spectrometers (GIS) were operated in the normal PH mode and the Solid-State Imaging Spectrometers (SIS) were operated in either FAINT or BRIGHT mode, usually with two CCD chips active. Data were screened and discarded when the X-ray telescopes (XRTs) were less than 10$^{\circ}$ from the Earth, the geomagnetic rigidity was smaller than 6 GeV, and the XRT for the SIS-0 and SIS-1 were less that 40$^{\circ}$ to 20$^{\circ}$ respectively from the sunlit Earth. Background was acquired from the same exposures, usually from annuli surrounding the cluster.

Source counts were accumulated from 6.125' and 2.5' radii regions for
the GIS and SIS respectively.  The different size regions reflect the different spatial resolutions of the two instruments.  The source spectra were grouped until
each bin contained at least 20 counts, the background was grouped with the
same binning and subtracted, and the net spectra from the four
detectors were simultaneously fitted with {\em XSPEC} 10. The adjustable
parameters were the GIS and SIS normalizations, temperature,
abundance, and hydrogen column density for the SIS. The fixed
parameters were the redshift and hydrogen column density for the
GIS. The total flux and luminosity were determined from the GIS
normalization, assuming the clusters are point sources. Our
temperature and luminosity measurements were usually in excellent
agreement with the values reported by others.

We used two different cosmologies for our analysis: the canonical matter-dominated cosmology with the present ratio of the density of matter in the universe to the critical density required to close the universe, $\Omega_{M}$, equal to one and $\Omega_{\Lambda} \equiv \frac{\Lambda}{3H_0^2}=0$ where $\Lambda$ is the cosmological constant and H$_0$ is the present Hubble constant.  The other is a cosmology with $\Omega_{M}$=0.3 and $\Omega_{\Lambda}$=0.7, which is consistent with the results from high redshift type Ia supernovae (Perlmutter \etal 1999, Schmidt \etal 1998) and many CMB results (e.g. Melchiorri \etal 2000).  We converted the X-Ray band luminosities into bolometric luminosities using bolometric corrections derived with {\em XSPEC}.   The properties of the high redshift clusters are listed in Table 1, where we include the luminosities and their errors for both cosmologies.  The mean redshift of the high redshift clusters is 0.34.  

For the z$<$0.3 clusters, we used the luminosities as published by White, Jones, \& Forman (1997).  White (2000) used {\em ASCA} data to determine the integrated non-cooling flow corrected temperatures of the clusters in the low redshift portion of our sample, and we use these values in our analysis.  We have listed these clusters and their properties in Table 2.  The published luminosities are for $\Omega_{M}$=1.0 and $\Omega_{\Lambda}$=0.0, which we converted to luminosities for the $\Omega_{M}$=0.3 $\Omega_{\Lambda}$=0.7 cosmology by using the ratio of their luminosity distances:
\begin{equation}
L_{i} = \frac{d_{L_i}^{2}}{d_{L_j}^{2}} L_{j}
\end{equation}
\begin{equation}
d_{L} = (1+z)\int_{0}^{z}\frac{c}{H_{0}}\mathrm{d}z^{\prime}[\Omega_{\Lambda} + \Omega_{M}(1+z^{\prime})^{3}]^{-0.5}
\end{equation}
The mean redshift of the low redshift clusters is 0.05; the mean redshift of all the clusters in the dataset is 0.179.  

\section{Analysis}

We devised a number of tests to examine how galaxy cluster luminosity scales with temperature, and whether or not the L-T relation has any redshift dependence. We assumed that the luminosity goes as a power law in temperature, as well as a power law in (1+z).  There is some intrinsic scatter in this relation.  Many authors have used the BCES method (Akritas \& Bershady,1996), which includes errors on all variables and allows for intrinsic scatter, to measure the slopes and scatter in the same relations (Markevitch 1998, \& Ikebe \etal 2001).  Our method from {\em Numerical Recipes} seems equivalent as we obtained nearly identical results for the temperature exponent.

We adoped the following form for the L-T relation: 
\begin{equation}
L_{bol,44} = C*T^{\alpha}*(1+z)^A. 
\end{equation}
Taking the logarithm of this equation, we found that:
\begin{equation}
log(L_{bol,44}) = log(C) + \alpha log(T) + Alog(1+z).
\end{equation}
For our analysis of these data, we included the errors on both the luminosity and the temperature, and we assumed that our redshifts do not contribute any error.  The discussion in {\it Numerical Recipes} (Press \etal 1988) describes how $\chi^{2}$ is related to the errors on each of the quantities, $\epsilon_L=\sigma$(logL)=0.4343$\sigma_L/L$ and $\epsilon_T=\sigma$(logT)=0.4343$\sigma_T/T$, where $\sigma_L$ and $\sigma_T$ are the measured errors on the luminosity and temperature.  The temperature errors were symmeterized by using the average of the plus and minus errors.  This discussion lead us to the following expression for $\chi^{2}$:
\begin{equation}
\chi^{2} = \sum_{i=1}^{N}\frac{(log(L_{bol,44,i}) - log(C) - \alpha log(T_i) - Alog(1+z_i))^{2}}{\epsilon_{L_i}^2 + \alpha^2\epsilon_{T_i}^2} 
\end{equation}
where N is the number of clusters.

By setting the derivative of $\chi^{2}$ with respect to log(C) equal to zero, the value of log(C) at the minimum $\chi^{2}$ can be expressed as a function of the other two parameters, A and $\alpha$:
\begin{equation}
log(C) = \frac{\sum_{i=1}^{N}(log(L_{bol,44,i})- \alpha log(T_i)- Alog(1+z_i))(\epsilon_{L_i}^2 + \alpha^2\epsilon_{T_i}^2)^{-1}}{\sum_{i=1}^{N}(\epsilon_{L_i}^2 + \alpha^2\epsilon_{T_i}^2)^{-1}}
\end{equation}
Substituting equation (6) into equation (5), we obtained an expression for $\chi^{2}$ in terms of A and $\alpha$.  We performed a grid search of $\chi^{2}$ to found the minimum with respect to A and $\alpha$ and solved for log(C) for the particular dataset.  The results for each dataset are given in Tables 3, 4, and 5.  We found the best fit values for each of the parameters for the two cosmologies for two different scenarios.  In one scenario we solved for A, and in the other scenario we set A=0 and fit for only $\alpha$.  In this way we can look at the L-T relation independent of the redshift of the galaxy clusters.  Table 3 contains the results of these calculations for the combined sample and Tables 4 and 5 contain the results we obtained  with the low and high redshift samples individually.  Figures 1-3 show a plot of our data with the best fits overlaid for both the A=0 case and the case where we fit for A.  Figures 2 and 3 show the best fits to the low or high redshift data as well as the best fit found using the combined sample for comparison.

\section{Error Analysis}
\subsection{Errors on the Parameters}
As there is intrinsic scatter in the data (see section 4.2 below), $\chi^{2}$ is quite large, so we resorted to numerical methods to estimate the error on each of the parameters.  We simulated 10,000 datasets using the bootstrap method.  The random datasets were created by randomly drawing N data points from the original dataset.  After each random draw, we put that point back into the original dataset so that it could possibly be drawn again.  This scenario is likely to produce a dataset which has a few clusters included more than once and others which may be excluded.  We then ran our $\chi^{2}$ minimization routine on each of the random datasets to fit our parameters.  To estimate 1-$\sigma$ errors on the parameters, we found the 16\% and 84\% confidence levels for each from the corresponding points of the cumulative distributions of each parameter as shown in figures 4-6.  These errors are included in Tables 3, 4 and 5 and are for the case of one interesting parameter.  In figures 7-9 we plot A vs. $\alpha$ for each of our Montecarlo simulations.  The ellipses shown are the smallest area encompassing 68\% and 95\% of the points, found using a grid search algorithm.  They are the errors for the case of two interesting parameters, also listed in Tables 3, 4, and 5.

In summary, we found that for $\Omega_{M}$=1.0, $\Omega_{\Lambda}$=0.0:  
\begin{equation}
log(L_{bol,44}) = (2.815^{+0.322}_{-0.316}\pm0.42)log(T) + (1.134^{+1.057}_{-1.073}\pm1.66)log(1+z) +(-1.167^{+0.216}_{-0.221}\pm0.25),
\end{equation}
and that for $\Omega_{M}$=0.3, $\Omega_{\Lambda}$=0.7:
\begin{equation}
log(L_{bol,44}) = (2.822^{+0.320}_{-0.323}\pm0.43)log(T) + (2.052^{+1.073}_{-1.058}\pm1.63)log(1+z) +(-1.126^{+0.223}_{-0.219}\pm0.26).
\end{equation}
where the errors are 68\% confidence for one or two interesting parameters respectively.

\subsection{Characterization of the Intrinsic Scatter in the L-T Relation}
To estimate the dispersion in the L-T relation, we compared our luminosity data with the calculated luminosities based on the temperatures of the clusters and equation 7 or 8.  We calculated two quantities: the difference between the predicted and measured luminosity in linear space ($\Delta$L) and in log space ($\Delta$logL).  The mean and standard deviation of these differences for $\Delta$L and $\Delta$logL for each of our sets of parameters are listed in Tables 6 and 7.  The distribution of $\Delta$L is non-gaussian, as shown in Figure 10, while the distribution of $\Delta$logL is approximately gaussian in nature as shown in Figure 11.  Over each of the histograms, we plotted a lognormal distribution:
\begin{equation}
f(x;\mu,\sigma)=(stepsize)*(N)[\frac{1}{\sigma\sqrt{2\pi}}exp{\frac{-(x-\mu)^2}{2\sigma^2}}]
\end{equation}
where x is either $\Delta$L (Figure 10) or $\Delta$logL (Figure 11), and $\mu$ and $\sigma$ are the corresponding mean and standard deviation.  The stepsize is the size of the bins in the histogram.  For the latter case, the distribution is lognormal.

The standard deviations in $\Delta$logL are very similar for all cases.  For A non-zero, we measured a standard deviation of 0.282 for $\Omega_{M}$=1.0, $\Omega_{\Lambda}$=0.0, and 0.283 for $\Omega_{M}$=0.3, $\Omega_{\Lambda}$=0.7. Markevitch (1998) measured a $\Delta$logL=0.181 for a sample of clusters with {\em ASCA} and {\em ROSAT} data without cooling flow corrections, which is somewhat lower than our measurements.  Ikebe \etal (2001), using a flux limited sample of galaxy clusters with {\em ROSAT} and {\em ASCA} data and correcting for cooling flows, measured $\Delta$logL=0.24.  This $\Delta$logL was measured with their $\alpha$ value of 2.47 (significantly lower than our average value) and including galaxy clusters down to a temperature of 1.4keV.  

\section{Conclusions}

Previous studies at both low and high redshift have found $\alpha\simeq$3 (see earlier references).  We found $\alpha$=3 within 1.52$\sigma$ for all of our data, independent of cosmology.  Borgani \etal (1999), using the X-Ray Luminosity Function (XLF) from the Rosat Deep Cluster Survey (Rosati \etal 1998) and the Brightest Cluster Survey (Ebeling \etal 1997), constrained $\alpha$ to be between 3 and 4.  Using their XLF and constraints on $\alpha$, they found 1$\le$A$\le$3 for an $\Omega_{M}$=1.0 universe and A=0 implied a low density universe.  Mushotzky \& Scharf (1997) found A=0.  Donahue \etal (1999) found a slightly negative value of A (for $\Omega_{M}$=1.0, A=$-$1.4$^{+0.8}_{-1.6}$, and for $\Omega_{M}$=0.3 $\Omega_{\Lambda}$=0.7, A=$-$0.8$^{+0.9}_{-1.1}$, as read from their figure 7), and rule out A=1.5; their results were also consistent with A=0.  Reichart, Castander and Nichol (1999) determined the L-T relation for cooling flow corrected luminosities and temperatures. They found $\alpha$ = 2.80$\pm$0.15 and A = 0.35$^{+0.54}_{-1.22}$ or 1.53$^{+0.54}_{-1.22}$ for $\Omega_M$=1, $\Omega_{\Lambda}$=0 or $\Omega_M$=0.3, $\Omega_{\Lambda}$=0.7 respectively. These results agree with ours.  Horner (2001) found for L$_{bol}$$>$2x10$^{44}$erg s$^{-1}$ that $\alpha$=2.98$\pm$0.14 and A=0.02$\pm$0.16 for no cooling flow corretions and $\Omega_M$=1.0 (his equation 5.7), again consistent with our results.

The A parameter is closely related to the duration and heating epoch of the ICM in the non-gravitational heating models (Cavaliere \etal 1997).  Bialek \etal (2001) found that a model using nonzero initial entropy yielded an $\alpha$ consistent with observations.  A consequence of this model is that A is approximately zero out to a redshift of 0.5.  Tozzi \& Norman (2001) included cooling and shocks in their model with an initial entropy, and they predicted A to be zero or only slightly positive out to a redshift of one.  They found that changing cosmology does not have a strong effect on the value of A.  The evolution expected in the cooling/drop out models has not yet been determined.  Obviously if the evolution is different from the heating models then we may be able to discriminate between them.

Our results indicate that the value of A for the matter-dominated $\Omega_{M}$=1.0 cosmology is smaller than for the $\Omega_{M}$=0.3 $\Omega_{\Lambda}$=0.7 case, but that both results are consistent with zero.  We conclude that our results are consistent with weak or no evolution of the L-T relation, and agree with the heating models mentioned above.

\acknowledgments
This work was supported by NASA grant NAG5-9166.  M. Sornig thanks the Institute for Astronomy for its hospitality during a visiting studentship.

\clearpage

\begin{deluxetable}{lcrccrcrc}
\tablecaption{Clusters with z$>$0.3} 
\tablehead{
\colhead{Cluster} & \colhead{Redshift} & \colhead{kT(keV)} & 
\colhead{+$\sigma_{T}$} & \colhead{$-\sigma_{T}$} & \colhead{L$_{bol,44}$} & 
\colhead{$\sigma_{L_{bol,44}}$} & \colhead{L$_{bol,44}$} & \colhead{$\sigma_{L_{bol,44}}$} \\
\colhead{} & \colhead{} & \colhead{} & 
(keV) & (kev) & \multicolumn{2}{c}{$\Omega_{M}$=1.0 $\Omega_{\Lambda}$=0.0}&
\multicolumn{2}{c}{$\Omega_{M}$=0.3 $\Omega_{\Lambda}$=0.7} \\
}
\startdata
A370      &  0.3730   &  7.04   &  0.520  &   0.500  & 28.72  & 0.631 &  38.685 & 0.850\\
A1722     &  0.3280   &  6.22   &  0.280  &   0.470  & 21.96  & 0.380 &  28.770 & 0.498\\
AC118     &  0.3080   &  10.61  &  0.490  &   0.480  & 62.16  & 0.524 &  80.389 & 0.678\\
A1995     &  0.3120   &  10.37  &  1.350  &   1.190  & 30.20  & 0.633 &  39.159 & 0.821\\
AC114     &  0.3120   &  8.81   &  0.620  &   0.610  & 31.74  & 0.532 &  41.155 & 0.690\\
A1576     &  0.3020   &  8.65   &  0.670  &   0.660  & 29.80  & 0.510 &  38.387 & 0.658\\
A959      &  0.3530   &  7.53   &  0.900  &   0.750  & 22.10  & 0.609 &  29.410 & 0.810\\
CL2244    &  0.3280   &  7.12   &  1.680  &   1.310  & 7.68   & 0.442 &  10.061 & 0.579\\
CL0500    &  0.3160   &  5.56   &  2.090  &   1.250  & 4.39   & 0.194 &   5.707 & 0.252\\
3C295     &  0.4599   &  7.29   &  1.640  &   0.440  & 25.64  & 0.841 &  36.264 & 1.189\\
A851      &  0.4069   &  8.52   &  0.970  &   0.920  & 18.35  & 0.488 &  25.210 & 0.670\\
CL2236    &  0.5520   &  5.51   &  1.610  &   1.120  & 10.40  & 0.904 &  15.396 & 1.339\\
RXJ1347   &  0.4510   &  11.16  &  0.490  &   0.480  & 197.52 & 1.815 & 278.043 & 2.554\\
MS2137.5  &  0.3130   &  4.85   &  0.290  &   0.290  & 33.98  & 0.711 &  44.089 & 0.922\\
MS1358.5  &  0.3290   &  6.92   &  0.510  &   0.490  & 21.36  & 0.652 &  28.001 & 0.855\\
MS0353.6  &  0.3200   &  6.46   &  0.980  &   0.800  & 14.31  & 0.528 &  18.651 & 0.688\\
MS1008.1  &  0.3060   &  8.21   &  1.150  &   1.050  & 18.30  & 0.592 &  23.635 & 0.765\\
MS1224.7  &  0.3260   &  4.09   &  0.650  &   0.520  & 7.09   & 0.403 &   9.277 & 0.527\\
MS1512.4  &  0.3730   &  3.39   &  0.400  &   0.350  & 9.59   & 0.606 &  12.917 & 0.816\\
MS1426.4  &  0.3200   &  6.38   &  0.980  &   1.200  & 9.62   & 0.422 &  12.539 & 0.550\\
MS1147.3  &  0.3030   &  5.96   &  0.990  &   0.690  & 8.68   & 0.318 &  11.189 & 0.409\\
MS0811.6  &  0.3120   &  4.87   &  0.950  &   0.630  & 5.69   & 0.333 &   7.378 & 0.432\\
MS1241.5  &  0.5490   &  6.09   &  1.380  &   1.140  & 22.59  & 1.270 &  33.396 & 1.878\\
MS0451.6  &  0.5392   &  10.27  &  0.850  &   0.800  & 66.51  & 1.239 &  97.873 & 1.823\\
MS0015.9  &  0.5466   &  8.92   &  0.570  &   0.560  & 72.62  & 1.237 & 107.237 & 1.827\\
MS0302.7  &  0.4246   &  4.35   &  0.800  &   0.640  & 11.18  & 0.741 &  15.513 & 1.028\\
MS1621.5  &  0.4274   &  6.59   &  0.920  &   0.810  & 15.93  & 0.612 &  22.138 & 0.850\\
MS2053.7  &  0.5830   &  8.14   &  3.680  &   2.150  & 13.88  & 1.100 &  20.840 & 1.652\\
CL0024    &  0.3901   &  5.11   &  1.210  &   0.950  & 5.80   & 0.637 &   7.892 & 0.867\\
MS0418.3  &  0.3500   &  7.02   &  1.350  &   1.390  & 5.60   & 0.253 &   7.438 & 0.336\\
MS0821.5  &  0.3470   &  5.90   &  1.560  &   1.470  & 3.26   & 0.277 &   4.322 & 0.367\\
MS1532.5  &  0.3200   &  4.12   &  0.710  &   0.570  & 3.89   & 0.258 &   5.070 & 0.336\\
\enddata
\end{deluxetable}

\clearpage

\renewcommand{\arraystretch}{0.8}
\begin{deluxetable}{lcrccrcrc}
\tablecaption{Clusters with z$<$0.3}
\tablehead{
\colhead{Cluster} & \colhead{Redshift} & \colhead{kT(keV)} & 
\colhead{+$\sigma_{T}$} & \colhead{$-\sigma_{T}$} & \colhead{L$_{bol,44}$} & 
\colhead{$\sigma_{L_{bol,44}}$} & \colhead{L$_{bol,44}$} & \colhead{$\sigma_{L_{bol,44}}$} \\
\colhead{} & \colhead{} & \colhead{} & 
(keV) & (kev) & \multicolumn{2}{c}{$\Omega_{M}$=1.0 $\Omega_{\Lambda}$=0.0}&
\multicolumn{2}{c}{$\Omega_{M}$=0.3 $\Omega_{\Lambda}$=0.7} \\
}
\startdata
  A85   &  0.0521 &  5.92  &   0.11  &	 0.11 & 16.511 & 0.098  & 17.401 & 0.103 \\   
  A119  &  0.0443 &  5.62  &   0.12  &	 0.12 & 5.781  & 0.147  &  6.047 & 0.154 \\   
  A262  &  0.0163 &  2.21  &   0.03  &	 0.03 & 0.915  & 0.030  &  0.931 & 0.031 \\ 
  A399  &  0.0715 &  6.80  &   0.17  &	 0.17 & 8.774  & 0.147  &  9.419 & 0.158 \\ 
  A400  &  0.0238 &  2.26  &   0.04  &	 0.04 & 0.756  & 0.065  &  0.775 & 0.067 \\ 
  A401  &  0.0748 &  8.68  &   0.17  &	 0.17 & 19.317 & 0.214  & 20.802 & 0.230 \\ 
  A426  &  0.0179 &  5.28  &   0.03  &	 0.03 & 21.434 & 1.031  & 21.835 & 1.050 \\ 
  A478  &  0.0881 &  6.58  &   0.26  &	 0.25 & 47.909 & 1.126  & 52.230 & 1.228 \\ 
  A496  &  0.0330 &  4.08  &   0.04  &	 0.04 & 5.808  & 0.212  &  6.007 & 0.219 \\ 
  A539  &  0.0288 &  2.89  &   0.05  &	 0.05 & 1.299  & 0.028  &  1.338 & 0.029 \\ 
 A548S  &  0.0415 &  3.10  &   0.10  &	 0.10 & 1.174  & 0.026  &  1.225 & 0.027 \\ 
  A576  &  0.0381 &  4.02  &   0.07  &	 0.07 & 2.761  & 0.096  &  2.870 & 0.100 \\ 
  A644  &  0.0704 &  7.47  &   0.12  &	 0.10 & 14.652 & 0.148  & 15.714 & 0.159 \\ 
  A665  &  0.1816 &  7.73  &   0.41  &	 0.35 & 25.395 & 0.719  & 29.980 & 0.849 \\ 
  A754  &  0.0542 &  9.83  &   0.27  &	 0.27 & 23.125 & 0.491  & 24.421 & 0.519 \\ 
 A1060  &  0.0124 &  3.19  &   0.03  &	 0.03 & 0.781  & 0.018  &  0.791 & 0.018 \\ 
 A1367  &  0.0214 &  3.64  &   0.10  &	 0.10 & 1.760  & 0.052  &  1.799 & 0.053 \\ 
 A1413  &  0.1427 &  7.32  &   0.26  &	 0.24 & 23.623 & 1.060  & 27.016 & 1.212 \\ 
 A1650  &  0.0840 &  6.11  &   0.15  &	 0.14 & 13.297 & 0.306  & 14.442 & 0.332 \\ 
 A1656  &  0.0231 &  8.67  &   0.17  &	 0.17 & 15.563 & 0.830  & 15.938 & 0.850 \\ 
 A1689  &  0.1810 &  9.23  &   0.28  &	 0.28 & 50.456 & 1.056  & 59.536 & 1.246 \\ 
 A1775  &  0.0696 &  3.69  &   0.20  &	 0.11 & 4.683  & 0.149  &  5.019 & 0.160 \\ 
 A1795  &  0.0621 &  5.80  &   0.07  &	 0.07 & 20.146 & 0.292  & 21.437 & 0.311 \\ 
 A2052  &  0.0348 &  3.03  &   0.04  &	 0.03 & 3.645  & 0.119  &  3.777 & 0.123 \\ 
 A2063  &  0.0355 &  3.52  &   0.09  &	 0.09 & 3.642  & 0.066  &  3.776 & 0.068 \\ 
 A2065  &  0.0722 &  5.42  &   0.13  &	 0.13 & 11.283 & 0.309  & 12.121 & 0.332 \\ 
 A2107  &  0.0421 &  3.93  &   0.10  &	 0.10 & 2.714  & 0.096  &  2.833 & 0.100 \\ 
 A2142  &  0.0899 &  9.02  &   0.32  &	 0.32 & 58.593 & 0.736  & 63.983 & 0.804 \\ 
 A2147  &  0.0356 &  4.50  &   0.07  &	 0.07 & 4.231  & 0.232  &  4.388 & 0.241 \\ 
 A2199  &  0.0299 &  4.27  &   0.04  &	 0.04 & 6.295  & 0.075  &  6.491 & 0.077 \\   
 A2218  &  0.1710 &  6.84  &   0.34  &	 0.04 & 18.093 & 0.750  & 21.179 & 0.878 \\   
 A2255  &  0.0808 &  6.87  &   0.20  &	 0.20 & 9.721  & 0.127  & 10.527 & 0.138 \\   
 A2256  &  0.0581 &  6.96  &   0.11  &	 0.11 & 15.777 & 0.115  & 16.724 & 0.122 \\   
 A2319  &  0.0559 &  9.73  &   0.27  &	 0.27 & 28.515 & 0.231  & 30.163 & 0.244 \\   
 A2634  &  0.0309 &  3.27  &   0.07  &	 0.07 & 1.204  & 0.022  &  1.243 & 0.023 \\   
 A2657  &  0.0414 &  3.81  &   0.07  &	 0.07 & 2.932  & 0.040  &  3.058 & 0.042 \\   
 A2670  &  0.0759 &  3.73  &   0.17  &	 0.13 & 3.885  & 0.178  &  4.188 & 0.192 \\   
 A3112  &  0.0746 &  4.45  &   0.07  &	 0.07 & 11.596 & 0.765  & 12.485 & 0.824 \\   
 A3158  &  0.0575 &  5.77  &   0.10  &	 0.05 & 6.970  & 0.137  &  7.384 & 0.145 \\   
 A3266  &  0.0594 &  8.34  &   0.17  &	 0.16 & 16.270 & 0.148  & 17.268 & 0.157 \\   
 A3391  &  0.0531 &  5.60  &   0.17  &	 0.17 & 3.372  & 0.071  &  3.557 & 0.075 \\   
 A3562  &  0.0499 &  3.42  &   0.16  &	 0.16 & 9.908  & 0.421  & 10.420 & 0.443 \\   
 A3571  &  0.0390 &  7.24  &   0.09  &	 0.09 & 17.157 & 0.238  & 17.852 & 0.248 \\   
 A3667  &  0.0542 &  7.13  &   0.14  &	 0.14 & 12.574 & 0.139  & 13.279 & 0.147 \\   
 A4059  &  0.0478 &  3.90  &   0.07  &	 0.07 & 4.936  & 0.179  &  5.180 & 0.188 \\   
  AWM4  &  0.0424 &  2.38  &   0.06  &	 0.06 & 1.673  & 0.022  &  1.747 & 0.023 \\   
  AWM7  &  0.0176 &  3.79  &   0.05  &	 0.05 & 3.119  & 0.041  &  3.176 & 0.042 \\   
 CYG-A  &  0.0570 &  9.49  &   0.23  &	 0.23 & 10.685 & 0.153  & 11.314 & 0.162 \\   
 MKW3S  &  0.0434 &  3.41  &   0.05  &	 0.05 & 4.259  & 0.110  &  4.451 & 0.115 \\   
  MKW4  &  0.0196 &  1.93  &   0.09  &	 0.08 & 0.445  & 0.007  &  0.454 & 0.007 \\   
  OPH   &  0.0280 &  12.75 &   0.29  &	 0.29 & 31.410 & 2.761  & 32.326 & 2.842 \\   
PKS0745 &  0.1028 &  7.21  &   0.11  &	 0.11 & 51.175 & 4.360  & 56.537 & 4.817 \\   
 2A0336 &  0.0350 &  3.08  &   0.03  &	 0.03 & 7.067  & 0.156  &  7.324 & 0.162 \\   
\enddata
\end{deluxetable}

\clearpage

\renewcommand{\arraystretch}{1}

\renewcommand{\arraystretch}{0.8}
\renewcommand{\tabcolsep}{1.1mm}
\begin{deluxetable}{ccccccccccccc}
\tablecaption{Combined Dataset Results} 
\tablehead{
\multicolumn{2}{c}{A} & \colhead{68\%} & \colhead{68\%} & 
\multicolumn{2}{c}{$\alpha$} & \colhead{68\%} & \colhead{68\%} & 
\multicolumn{2}{c}{log(C)} & \colhead{68\%} & \colhead{68\%} & \colhead{Cosmology} \\
\colhead{c\footnote[1]}& \colhead{mc\footnote[2]} & \colhead{1-par} &  \colhead{2-par} & 
\colhead{c\footnote[1]}& \colhead{mc\footnote[2]} & \colhead{1-par} &  \colhead{2-par}  & 
\colhead{c\footnote[1]}& \colhead{mc\footnote[2]} & \colhead{1-par} &  \colhead{2-par}  & \colhead{}\\
}
\startdata
1.134 & 1.146 & $^{+1.057}_{-1.073}$ & $\pm1.66$ & 2.815 & 2.855 & $^{+0.322}_{-0.316}$ & $\pm0.42$ & $-$1.167 & $-$1.167 & $^{+0.216}_{-0.221}$ & $\pm0.25$& {\scriptsize $\Omega_{M}$=1.0 $\Omega_{\Lambda}$=0.0} \\
0     &  $\cdots$ &  $\cdots$  &  $\cdots$ & 2.910 & 2.948 & $^{+0.283}_{-0.289}$ &  $\cdots$ & $-$1.162 & $-$1.203 & $^{+0.213}_{-0.208}$ &  $\cdots$ & {\scriptsize $\Omega_{M}$=1.0 $\Omega_{\Lambda}$=0.0} \\
2.052 & 2.065 & $^{+1.073}_{-1.058}$ & $\pm1.63$ & 2.822 & 2.862 & $^{+0.320}_{-0.323}$ & $\pm0.43$ & $-$1.126 & $-$1.169 & $^{+0.223}_{-0.219}$ & $\pm0.26$ & {\scriptsize $\Omega_{M}$=0.3 $\Omega_{\Lambda}$=0.7} \\ 
0     &   $\cdots$ &  $\cdots$ &  $\cdots$ & 3.003 & 3.040 & $^{+0.290}_{-0.297}$ &  $\cdots$ & $-$1.201 & $-$1.242 & $^{+0.220}_{-0.215}$ &  $\cdots$ & {\scriptsize $\Omega_{M}$=0.3 $\Omega_{\Lambda}$=0.7} \\ 
\enddata
\tablenotetext{1}{c = the $\chi^2$ minimum.}
\tablenotetext{2}{mc = the mean of the Montecarlo parameter estimates.}
\end{deluxetable}

\renewcommand{\arraystretch}{1}
\renewcommand{\tabcolsep}{2mm}

\clearpage

\renewcommand{\arraystretch}{.8}
\renewcommand{\tabcolsep}{1.1mm}

\begin{deluxetable}{cccccccccccccc}
\tablecaption{Individual Dataset Results:  z$<0.3$}
\tablehead{
\multicolumn{2}{c}{A} & \colhead{68\%} & \colhead{68\%} & 
\multicolumn{2}{c}{$\alpha$} & \colhead{68\%} & \colhead{68\%} & 
\multicolumn{2}{c}{log(C)} & \colhead{68\%} & \colhead{68\%} & \colhead{Cosmology} \\
\colhead{c\footnote[1]}& \colhead{mc\footnote[2]} & \colhead{1-par} &  \colhead{2-par} & 
\colhead{c\footnote[1]}& \colhead{mc\footnote[2]} & \colhead{1-par} &  \colhead{2-par}  & 
\colhead{c\footnote[1]}& \colhead{mc\footnote[2]} & \colhead{1-par} &  \colhead{2-par}  & \colhead{}\\
}
\startdata
0.021 & 0.176 & $^{+3.584}_{-3.572}$ & $\pm5.47$ & 2.841 & 2.863 & $^{+0.354}_{-0.353}$ & $\pm0.49$ & $-$1.120 & $-$1.154 & $^{+0.225}_{-0.225}$ &  $\pm0.26$ & {\scriptsize $\Omega_{M}$=1.0 $\Omega_{\Lambda}$=0.0} \\
    0 &   $\cdots$ &  $\cdots$  &  $\cdots$ & 2.842 & 2.864 & $^{+0.305}_{-0.307}$  &  $\cdots$ & $-$1.120 & $-$1.151 & $^{+0.224}_{-0.220}$ &  $\cdots$ & {\scriptsize  $\Omega_{M}$=1.0 $\Omega_{\Lambda}$=0.0} \\
1.012 & 1.044 & $^{+3.467}_{-3.700}$ & $\pm4.89$ & 2.845 & 2.867 & $^{+0.357}_{-0.348}$ & $\pm0.55$ & $-$1.122 & $-$1.156 & $^{+0.224}_{-0.228}$ & $\pm0.30$ & {\scriptsize $\Omega_{M}$=0.3 $\Omega_{\Lambda}$=0.7} \\ 
    0 &   $\cdots$ &  $\cdots$  &  $\cdots$ & 2.896 & 2.918 & $^{+0.315}_{-0.316}$  &  $\cdots$ & $-$1.136 & $-$1.167 & $^{+0.231}_{-0.227}$ &  $\cdots$ & {\scriptsize $\Omega_{M}$=0.3 $\Omega_{\Lambda}$=0.7} \\ 
\enddata
\tablenotetext{1}{c = the $\chi^2$ minimum.}
\tablenotetext{2}{mc = the mean of the Montecarlo parameter estimates.}
\end{deluxetable}

\renewcommand{\arraystretch}{1}
\renewcommand{\tabcolsep}{2mm}

\clearpage

\renewcommand{\arraystretch}{.8}
\renewcommand{\tabcolsep}{1.1mm}

\begin{deluxetable}{ccccccccccccc}
\tablecaption{\bf Individual Dataset Results: 0.3$<$z$<0.6$}
\tablehead{
\multicolumn{2}{c}{A} & \colhead{68\%} & \colhead{68\%} & 
\multicolumn{2}{c}{$\alpha$} & \colhead{68\%} & \colhead{68\%} & 
\multicolumn{2}{c}{log(C)} & \colhead{68\%} & \colhead{68\%} & \colhead{Cosmology} \\
\colhead{c\footnote[1]}& \colhead{mc\footnote[2]} & \colhead{1-par} &  \colhead{2-par} & 
\colhead{c\footnote[1]}& \colhead{mc\footnote[2]} & \colhead{1-par} &  \colhead{2-par}  & 
\colhead{c\footnote[1]}& \colhead{mc\footnote[2]} & \colhead{1-par} &  \colhead{2-par}  & \colhead{}\\
}
\startdata
1.385 & 0.251 & $^{+3.260}_{-3.634}$ & $\pm4.79$ & 3.809 & 4.082 & $^{+0.912}_{-0.655}$ & $\pm1.17$ & $-$2.023 & $-$2.097 & $^{+0.484}_{-0.618}$ & $\pm0.75$ & {\scriptsize   $\Omega_{M}$=1.0 $\Omega_{\Lambda}$=0.0} \\
0     &  $\cdots$  &  $\cdots$ &  $\cdots$  & 3.960 & 4.009 & $^{+0.683}_{-0.529}$  &  $\cdots$ & $-$1.967 & $-$2.015 & $^{+0.491}_{-0.587}$ &  $\cdots$ & {\scriptsize   $\Omega_{M}$=1.0 $\Omega_{\Lambda}$=0.0} \\
2.181 & 1.169 & $^{+3.382}_{-3.512}$ & $\pm4.76$ & 3.810 & 4.083 & $^{+0.914}_{-0.622}$ & $\pm1.19$ & $-$2.005 & $-$2.079 & $^{+0.470}_{-0.618}$ & $\pm0.84$ & {\scriptsize  $\Omega_{M}$=0.3 $\Omega_{\Lambda}$=0.7} \\ 
0     &  $\cdots$  &  $\cdots$ &  $\cdots$  & 4.068 & 4.108 & $^{+0.553}_{-0.695}$  &  $\cdots$ & $-$1.934 & $-$1.975 & $^{+0.596}_{-0.509}$ &  $\cdots$ & {\scriptsize  $\Omega_{M}$=0.3 $\Omega_{\Lambda}$=0.7} \\ 
\enddata
\tablenotetext{1}{c = the $\chi^2$ minimum.}
\tablenotetext{2}{mc = the mean of the Montecarlo parameter estimates.}
\end{deluxetable}

\renewcommand{\tabcolsep}{2mm}
\renewcommand{\arraystretch}{1.0}

\clearpage

\begin{deluxetable}{cccrcc}
\tablecaption{\bf Difference of $\Delta$L from the L-T Relation}
\tablehead{
\colhead{A} & \colhead{$\alpha$} & \colhead{log(C)} & \colhead{mean} & 
\colhead{standard deviation} & \colhead{Cosmology}
}
\startdata
1.134 & 2.815 & $-$1.167 & $-$1.034 & 19.66 & $\Omega_{M}$=1.0 $\Omega_{\Lambda}$=0.0 \\
0 & 2.910 &  $-$1.162 & 0.9185 & 19.74 & $\Omega_{M}$=1.0 $\Omega_{\Lambda}$=0.0 \\
2.053 & 2.822 & $-$1.126 & $-$5.182 & 27.77 &$\Omega_{M}$=0.3 $\Omega_{\Lambda}$=0.7 \\ 
0 & 3.003 & $-$1.202 & 2.745 & 28.29 &$\Omega_{M}$=0.3 $\Omega_{\Lambda}$=0.7 \\ 
\enddata
\end{deluxetable}

\clearpage

\begin{deluxetable}{cccrcc}
\tablecaption{\bf Difference of $\Delta$log(L) from the L-T Relation}
\tablehead{
\colhead{A} & \colhead{$\alpha$} & \colhead{log(C)} & \colhead{mean} & 
\colhead{standard deviation} & \colhead{Cosmology}
}
\startdata
1.134 & 2.815 & $-$1.167 & 0.0690 & 0.282 & $\Omega_{M}$=1.0 $\Omega_{\Lambda}$=0.0 \\
0 & 2.910 & $-$1.162 & 0.0254 & 0.285 & $\Omega_{M}$=1.0 $\Omega_{\Lambda}$=0.0 \\
2.053 & 2.822 & $-$1.126 & 0.0691 & 0.283 &$\Omega_{M}$=0.3 $\Omega_{\Lambda}$=0.7 \\ 
0 & 3.003 & $-$1.202 & $-$0.0083 & 0.298 &$\Omega_{M}$=0.3 $\Omega_{\Lambda}$=0.7 \\ 
\enddata
\end{deluxetable}

\end{document}